\documentclass{elsart}
\usepackage{epsf}

\usepackage{amssymb}

\begin{document}

\begin{frontmatter}



\title{Impurity Effect on Ferromagnetic Transition\\
in Double-Exchange Systems
}


\author[Motome]{Y. Motome\corauthref{cor}},
%\thanks{fax: +81-298-55-7440, E-mail: motome@ims.tsukuba.ac.jp}
\corauth[cor]{tel: +81-298-53-5028, fax: +81-298-55-7440}
\ead{motome@ims.tsukuba.ac.jp}
\author[Furukawa]{N. Furukawa}

\address[Motome]{Institute of Materials Science, University of Tsukuba,
1-1-1 Tennou-dai, Tsukuba, Ibaraki 305-0006, Japan}
\address[Furukawa]{Department of Physics, Aoyama Gakuin University, 
6-16-1 Chitose-dai, Setagaya, Tokyo 157-8572, Japan}

\begin{abstract}
Effect of randomness in the double-exchange model is studied.
Large fluctuations and spatial random distribution of impurities
are taken into account in an essentially exact manner
by using the Monte Carlo calculation.
The randomness suppresses the ferromagnetism
by reducing the coherence of itinerant electrons.
The suppression is significant in the critical region
where the fluctuations are dominant.
Temperature dependences of the magnetization are estimated
for finite-size clusters.
A characteristic temperature for phase transition
$T^{*}$ is estimated from the inflection point,
which is expected to give a good approximation for
the critical temperature in the thermodynamic limit.
Our results suggest that the ferromagnetism becomes unstable
more rapidly than predicted in the previous theoretical results
by the coherent-potential approximation.
\end{abstract}

\end{frontmatter}

% main text

%%%%%%%%%%%%%%%%%%%%%%%%%%%%%%%%%%%%%%%%%%%%%%%%%%%%%%%%%%%%%
\section{Introduction}
\label{Sec:Introduction}
%%%%%%%%%%%%%%%%%%%%%%%%%%%%%%%%%%%%%%%%%%%%%%%%%%%%%%%%%%%%%
The double-exchange (DE) model has been studied intensively
to explain ferromagnetic transition
in colossal-magnetoresistance manganites.
\cite{Furukawa1999}
The origin of the metallic ferromagnetism in these materials
is well explained by the DE mechanism.
\cite{Zener1951}
Recently, numerical calculations have demonstrated that
fluctuations through the strong interplay
between spin and charge degrees of freedom play a crucial role
in this system.
\cite{Motome2000,Motome2001a,Motome2001b,Furukawa2001a}
Through precise calculations of 
the ferromagnetic transition temperature $T_{\rm c}$,
the DE model is shown to give a quantitative description
of the ferromagnetic transition in the typical material
La$_{1-x}$Sr$_{x}$MnO$_{3}$ near $x=0.3$.
This provides a well-established starting point
toward comprehensive understanding of
rich variety of physics in the manganites
through quantitative comparisons between experiments and theories.

To this canonical DE ferromagnetic state,
there are various perturbations in experiments,
for instance, control of chemical compositions,
external pressure and magnetic field.
\cite{Ramirez1997}
It is important to clarify these effects theoretically
in order to understand experiments
where the perturbations may be entangled and act on in a complicated manner.
Among them, we study effects of the randomness in this paper.
Since mixed-valence manganites are synthesized as solid solutions,
randomness is inevitably contained in these compounds.
We discuss how the ferromagnetic transition is affected
by the randomness.

The randomness effect for $T_{\rm c}$ in the DE system
has been studied theoretically.
Stability of the ferromagnetic state is studied by
the coherent-potential approximation (CPA), and
a decrease of $T_{\rm c}$ due to the randomness is predicted.
\cite{Letfulov2001,Auslender2001a,Auslender2001b}
In this work, we go beyond the approximation and
use a numerical technique which is essentially exact.
Spatial random distribution of impurities is
taken into account properly in this method.
Since the spatial fluctuations which are neglected in the CPA
are known to be important in the case without disorder,
\cite{Motome2000}
the unbiased calculations are necessary also for the systems with disorder
to discuss the quantitative change of $T_{\rm c}$.

This paper is organized as follows.
The next section \ref{Sec:Model} introduces the DE model
with disorder and describes the numerical methods briefly.
In Sec.~\ref{Sec:Results}, the results for 
the approximately-estimated critical temperature
are discussed for the cases with randomness
in comparison with the results for the pure system.
Sec.~\ref{Sec:Summary} is devoted to summary and concluding remarks.

%%%%%%%%%%%%%%%%%%%%%%%%%%%%%%%%%%%%%%%%%%%%%%%%%%%%%%%%%%%%%
\section{Model and Method}
\label{Sec:Model}
%%%%%%%%%%%%%%%%%%%%%%%%%%%%%%%%%%%%%%%%%%%%%%%%%%%%%%%%%%%%%
In this work, we consider a simple model with a diagonal disorder
in the framework of the DE model.
The Hamiltonian is explicitly given by
\begin{equation}
H = - t \sum_{\langle ij \rangle, \sigma}
( c_{i \sigma}^{\dagger} c_{j \sigma} + {\rm h.c.} )
- J_{\rm H} \sum_{i} \vec{\sigma}_{i} \cdot \vec{S}_{i}
+ \sum_{i \sigma} \varepsilon_{i} c_{i \sigma}^{\dagger} c_{i \sigma},
\label{eq:H}
\end{equation}
where 
the first term denotes the nearest-neighbor hopping on the cubic lattice
and the second term is for the Hund's-rule coupling.
The last term denotes the on-site randomness
which modifies the one-body potential energy in each site.
The potential $\varepsilon_{i}$ takes $\pm W_{\rm imp}/2$
in equal probability in each site.
The physical quantities are calculated by the random average
for the quenched disorder.
For simplicity, 
we consider the limits of $J_{H} \rightarrow \infty$
and $|\vec{S}| \rightarrow \infty$.
In the following, we discuss the quarter-filled case
which shows a maximum of $T_{\rm c}$ in the pure case.

We apply the Monte Carlo (MC) method to the model (\ref{eq:H}).
\cite{Yunoki1998,Motome1999,Furukawa2001b}
In this method, the MC weight to update a configuration of
the localized moments are calculated quantum-mechanically
by integrating out the fermion degrees of freedom.
For large-size clusters, the difficulty of
a rapid increase of the computational time
has been solved by using the moment-expansion MC method.
\cite{Motome1999,Furukawa2001b}
The moment expansion as well as the random average is
performed quite efficiently on parallel computer systems.

In the following, we use an energy unit $W = 6t = 1$
which is the half bandwidth for the pure system.
For a given configuration of disorder,
we have typically run $1,000$ MC samplings for measurements
after $1,000$ MC steps for thermalization.
The random average is taken for $50$ random configurations.

%%%%%%%%%%%%%%%%%%%%%%%%%%%%%%%%%%%%%%%%%%%%%%%%%%%%%%%%%%%%%
\section{Results and Discussions}
\label{Sec:Results}
%%%%%%%%%%%%%%%%%%%%%%%%%%%%%%%%%%%%%%%%%%%%%%%%%%%%%%%%%%%%%
Before going to the cases with randomness,
we review and discuss the ferromagnetic transition
in the pure case ($W_{\rm imp} = 0$).
\cite{Motome2000}
Figure~\ref{fig:mvsT,pure} shows the results for
the temperature dependence of the magnetization for $W_{\rm imp}=0$.
We present the data for finite-size clusters and
those for the thermodynamic limit.
The latter is obtained by the finite-size scaling analysis.
From the onset of the magnetization in the thermodynamic limit,
$T_{\rm c}$ is estimated as $T_{\rm c}/W = 0.023 \pm 0.003$.

\begin{figure}
\epsfxsize=8cm
\centerline{\epsfbox{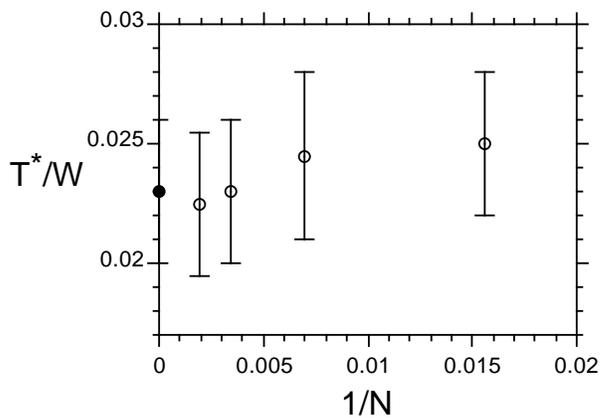}}
\caption{
System-size dependence of the temperature of the inflection points
in the magnetization curve for the finite-size clusters.
The data on the vertical axis is the estimate of $T_{\rm c}$
in the thermodynamic limit.
}
\label{fig:inflection}
\end{figure}

Figure~\ref{fig:mvsT,pure} indicates that
in order to obtain the precise estimate of $T_{\rm c}$,
a systematic study of the finite-size effects is crucial,
which needs much computational effort.
We note, however, that it is possible to define
a characteristic temperature even for a fixed-size cluster.
The temperature dependence of the magnetization
for each finite-size cluster is a monotonic function and
has an inflection point.
We adopt this inflection point
as the characteristic temperature $T^{*}$.
In Fig.~\ref{fig:inflection}, we plot the system-size dependence of $T^{*}$
which is obtained from the numerical differentiation of 
the data in Fig.~\ref{fig:mvsT,pure}.
Within the error bars,
$T^{*}$ is almost independent of the system size
and agrees with $T_{\rm c}$ in the thermodynamic limit.
Thus, $T^{*}$ for a finite-size cluster is expected to give
a good approximation for $T_{\rm c}$ in this pure case.
We expect that this observation holds also
for the systems with randomness below.
Note that at least, since $T^{*}$ must be an independent parameter
to characterize the temperature scale for the ferromagnetic transition,
it should be justified to discuss the stability of the ferromagnetism
by $T^{*}$.

Now we discuss the randomness effect.
Figure~\ref{fig:mvsT,disorder} shows the magnetization curve
for the system size of $4 \times 4 \times 4$ sites
when we change the value of $W_{\rm imp}$.
The magnetization decreases as the disorder strength $W_{\rm imp}$ increases.
This indicates that the ferromagnetism is suppressed
by the randomness.
The decrease is the most significant near $T_{\rm c}(W_{\rm imp}=0)$.
This is probably because near the critical point,
the fluctuations are dominant and
the coherence of the itinerant electrons
becomes very sensitive to the randomness.

Based on the discussion in the pure case,
we examine the stability of the ferromagnetism
in the systems with randomness by $T^{*}$.
Figure~\ref{fig:Tc,disorder} shows $T^{*}$
as a function of $W_{\rm imp}$
which is estimated from the data in Fig.~\ref{fig:mvsT,disorder}.
As the critical temperature in the previous CPA results,
\cite{Letfulov2001,Auslender2001a,Auslender2001b}
the approximated critical temperature $T^{*}$
decreases as $W_{\rm imp}$ increases.
Our MC data, however, show larger decrease
in comparison with the CPA results: For instance,
for $W_{\rm imp}/W = 0.5$, our data show about $35$\% decrease,
while the CPA results show less than $20$\% decrease.
The CPA takes account of the randomness
in an averaged way through the change of the density of states.
In our MC study, the spatial distribution of the randomness
as well as its effect on the coherence of electrons
is included properly.
Our data suggest that the spatial fluctuations are important and
the suppression of $T_{\rm c}$ is stronger than
expected from the band renormalization in the CPA.

%%%%%%%%%%%%%%%%%%%%%%%%%%%%%%%%%%%%%%%%%%%%%%%%%%%%%%%%%%%%%
\section{Summary and Concluding Remarks}
\label{Sec:Summary}
%%%%%%%%%%%%%%%%%%%%%%%%%%%%%%%%%%%%%%%%%%%%%%%%%%%%%%%%%%%%%
We have investigated the randomness effect
on the ferromagnetic transition in the double-exchange model.
We consider the model with on-site diagonal disorder.
The spatial distribution of the randomness and
large fluctuations inherent in this strongly-correlated system
are taken into account properly by using the Monte Carlo method.
The randomness suppresses the ferromagnetism
of the double-exchange origin
by reducing the coherence of the electron motion.
Our results indicate that the suppression is significant
in the critical region where the fluctuations are dominant.
We have examined the characteristic temperature scale
which is expected to give a good approximation for the critical temperature.
The approximated critical temperature 
tends to be suppressed by the randomness
as predicted by the coherent-potential approximation.
Our Monte Carlo results, however, show a more rapid decrease
than the previous results.
This may be attributed to the spatial fluctuations by the randomness distribution
which is included only in an averaged manner in the previous approximations.

The randomness effects are studied
only for a small-size cluster in this paper.
The estimated temperature $T^{*}$ is an approximation for $T_{\rm c}$.
Calculations for larger-size systems and
more precise determination of the critical temperature
including the finite-size scaling are under investigation.
Different distribution of the randomness is also examined
and will be reported elsewhere.

%%%% ACKNOWLEDGMENT %%%
\section*{Acknowledgement}

The authors thank H. Nakata for helpful support
in developing parallel-processing systems.
The computations have been performed mainly 
using the facilities in the AOYAMA+ project
(http://www.phys.aoyama.ac.jp/\~{}aoyama+)
and in the Supercomputer Center, Institute for Solid State Physics,
University of Tokyo.
This work is supported by  ``a Grant-in-Aid from
the Ministry of Education, Culture, Sports, Science and Technology''.

\newpage

%%%%% FIGURES %%%%%

\begin{figure}
\epsfxsize=8cm
\centerline{\epsfbox{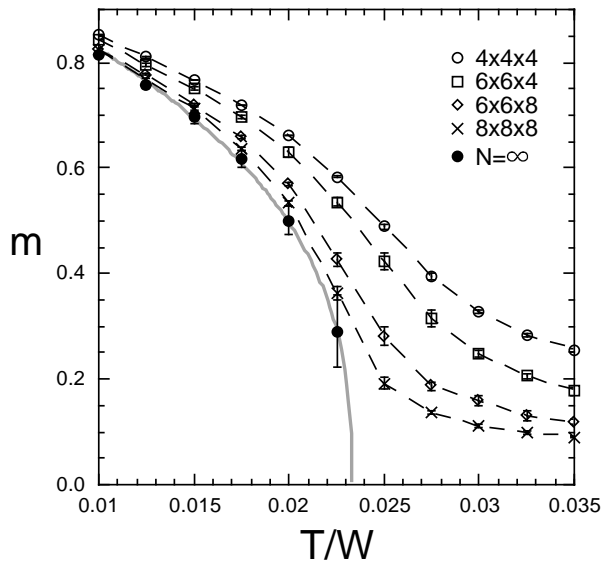}}
\caption{
Temperature dependence of the magnetization for the pure system.
The dashed lines connecting the data for finite-size clusters
are guides for the eye.
The gray curve for the data in the thermodynamic limit is
a scaling fit of $m \sim (T_{\rm c} - T)^{\beta}$.
}
\label{fig:mvsT,pure}
\end{figure}

\begin{figure}
\epsfxsize=8cm
\centerline{\epsfbox{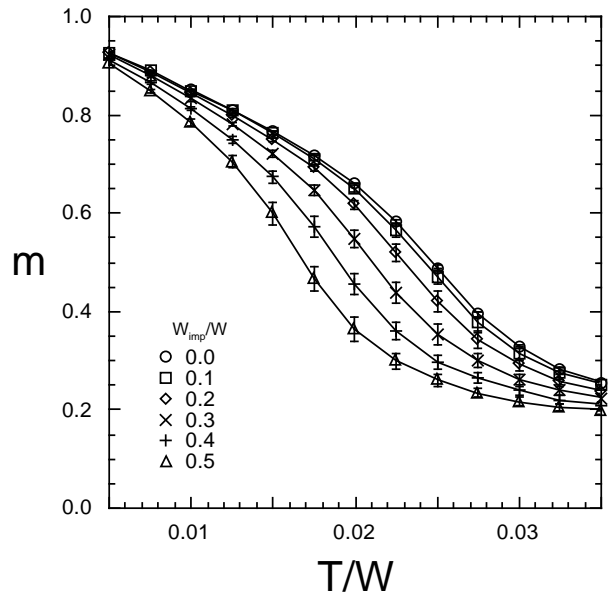}}
\caption{
Temperature dependence of the magnetization for the systems with randomness.
The data are for the cluster with $4 \times 4 \times 4$ sites.
The lines are guides for the eye.
}
\label{fig:mvsT,disorder}
\end{figure}

\begin{figure}
\epsfxsize=8cm
\centerline{\epsfbox{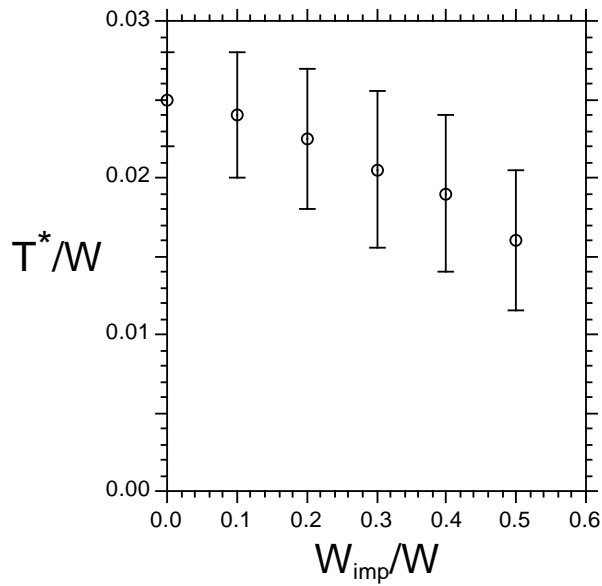}}
\caption{
The approximately-estimated transition temperature
as a function of the strength of the randomness.
The data are estimated from the inflection points
in the magnetization curve in Fig.~\ref{fig:mvsT,disorder}.
}
\label{fig:Tc,disorder}
\end{figure}

\end{document}